\documentclass[twocolumn]{aastex62}

\usepackage{amsmath}
\usepackage{graphicx}
\usepackage{natbib}
\usepackage[caption=false]{subfig}
\usepackage[nameinlink]{cleveref}

\newcommand{\kms}{km\ s$^{-1}$}
\newcommand{\kmskpc}{km\ s$^{-1}$\ kpc$^{-1}$}
\newcommand{\gaia}{\emph{Gaia}}
\newcommand{\VR}{$\tilde{V}_R$}
\newcommand{\VPHI}{$\tilde{V}_\phi$}
\newcommand{\dVPHI}{$\tilde{V}_\phi - V_{\rm LSR}$}
\newcommand{\VZ}{$\tilde{V}_Z$}
\newcommand{\EVR}{$\epsilon_{\tilde{V}_R}$}
\newcommand{\EVPHI}{$\epsilon_{\tilde{V}_\phi}$}
\newcommand{\EVZ}{$\epsilon_{\tilde{V}_Z}$}

\begin{document}

\title{Ripple patterns in in-plane velocities of OB stars from LAMOST and \gaia}

\author[0000-0002-7009-3957]{Xinlun Cheng}
\affil{Physics Department and Tsinghua Centre for Astrophysics, Tsinghua University, Beijing 100084, China}

\author[0000-0002-1802-6917]{Chao Liu}
\affil{Key Laboratory for Optical Astronomy, National Astronomical Observatories, Chinese Academy of Sciences, Beijing 100012, China}

\author[0000-0001-8317-2788]{Shude Mao}
\affil{Physics Department and Tsinghua Centre for Astrophysics, Tsinghua University, Beijing 100084, China}
\affil{National Astronomical Observatories, Chinese Academy of Sciences, 20A Datun Road, Chaoyang District, Beijing 100012, China}

\author[0000-0003-1359-9908]{Wenyuan Cui}
\affil{Department of Physics, Hebei Normal University, Shijiazhuang 050024, China}

\correspondingauthor{Chao Liu}
\email{liuchao@nao.cas.cn}

\begin{abstract}
    With about 12\,000 OB type stars selected from the LAMOST and \gaia\ survey, we study their 3 dimensional velocity distribution over the range of galactocentric radius from 6 to 15\,kpc in the Galactic disk plane. A clear ripple pattern in the radial velocity ($V_R$) map is shown. The median $V_R$ reaches $-8$ km\ s$^{-1}$ at $R\sim9$\,kpc, then increases to $\sim0$\,km\ s$^{-1}$ at $R\sim12$\,kpc, and later declines to below $-10$\,km\ s$^{-1}$ beyond $R\sim13$\,kpc. The median azimuthal velocity ($V_\phi$) map shows a similar pattern but has roughly $1/4$ phase difference with the radial velocity. Although the ripple of negative $V_R$ at $\sim9$\,kpc extends to about 40$^\circ$ in the azimuth angle, it does not align with either the Local or the Perseus spiral arm. Moreover, the farther ripple beyond 13\,kpc does not match the Outer spiral arm either. This indicates that the non-axisymmetric kinematic features are not induced by perturbations of known spiral structures. The central rotating bar can not lead to such patterns in the outer disk either. External perturbation of a dwarf galaxy or a dark matter sub-halo can induce such patterns but requires more evidence from both observations and simulations. The $V_\phi$ map in the $Z$--$V_Z$ plane of the OB stars is also investigated. Despite asymmetry to some degree, no spiral pattern is found. This is reasonable since most of the OB stars have ages much younger than 100 Myrs, which is smaller than one orbital period around the Galactic center.  
\end{abstract}


\section{Introduction}\label{sec:intro}
Following \gaia\ DR2, the phase-mixing features in the $Z$--$V_Z$ phase space of the Milky Way was discovered  \citep{2018arXiv180410196A} and confirmed \citep{2018arXiv180902658B,2018ApJ.865L.19T}. This is likely caused by the perturbation of a satellite galaxy passing through the Galactic disk. \citet{2018arXiv180410196A} suggests that the encounter with the dwarf galaxy happened between 300 and 900 Myrs ago, while \citet{2018ApJ.865L.19T} and \citet{2018arXiv180902658B} favor that the event occurred $\sim500$\,Myrs ago.

The possible effect of interaction with satellite galaxies has been explored a long time ago \citep{1997MNRAS.287..947E}. Since the discovery of the phase-mixing phenomenon, both toy-models and N-body simulations have been performed, and found that the external perturber may induce coupled effects in both vertical and in-plane directions \citep{2016ApJ...823....4D, 2018MNRAS.481.1501B, 2018arXiv180902658B}. 

In addition, non-axisymmetric features in the velocity and density distributions have also been explored. The radial velocity, $V_R$, asymmetry was firstly detected with stars from the Radial Velocity Experiment (RAVE) survey \citep{2011MNRAS.412.2026S}, later confirmed with various stellar tracers from different surveys~\citep{2013ApJ.777L.5C,2015RAA.15.1342S,2017RAA.17.114T, 2018IAUS.334.109L,2013MNRAS.436.101W,2018A&A.616A.11G}. Various theoretical explanations have been offered: perturbation by the central rotating bar \citep{2001A&A...373..511F, 2017RAA.17.114T,2018IAUS.334.109L}, spiral arms \citep{2012MNRAS.425.2335S, 2018A&A.616A.11G}, or a merging satellite galaxy \citep{2013ApJ.777L.5C}.

Vertical density and velocity oscillations have also been extensively studied (e.g., \cite{2012ApJ.750L.41W,2015ApJ.801.105X,2018MNRAS.478.3367W,2018arXiv180903507B, 2013ApJ.777L.5C, 2015RAA.15.1342S, 2018MNRAS.477.2858W}). Theoretical explanations have been proposed, including the interaction with the Sagittarius \citep{2013MNRAS.429.159G} and vertical perturbation by spiral structures \citep{1976MNRAS.177..265N, 2014MNRAS.443L.1D, 2014MNRAS.440.2564F}. 

It is noted that most previous studies are based on old stellar populations with age of at least $1$\,Gyr. Investigating asymmetric kinematics in young populations can play an important role since they can reflect the perturbed kinematics of the gaseous disk. In this work we use OB stars, which have ages from a few million to a few tens of millions years, to explore asymmetric motions in the 3 dimensional velocity map beyond 6\,kpc from the Sun in the Galactic plane.

The paper is organized as follows: the procedure for data selection and analysis is in Section~\ref{sec:data}, results and discussions are in Section~\ref{sec:res}, and conclusions are drawn in Section~\ref{sec:conc}.

\section{Data}\label{sec:data}
The data are combined with parameters from two surveys, line-of-sight velocities from LAMOST DR5 \citep{2012RAA.12.723Z, 2012RAA.12.735D, 2015RAA.15.1095L} and proper motions from the \gaia\ DR2 \citep{2016A&A.595A.1G, 2018A&A.616A.1G}. We employed around 16\,000 OB stars with spectral signal-to-noise ratio larger than 15 selected from LAMOST DR5 (Liu et al. submitted). The distance of these OB stars are derived by \citet{2018AJ.156.58B}. To keep the reliability of the distances, we only use OB stars with relative distance uncertainty $\sigma_d/d<0.4$, in which $d$ and $\sigma_d$ are distance and its uncertainty. In fact, most stars have $0.07<\sigma_d/d<0.15$, which is considerably lower than the cut-off value ($0.4$).

Radial velocity was derived from $H\delta$ and $H\gamma$ absorption lines of the stellar spectra by using a forward model based on TLUSTY synthetic spectra library \citep{2007ApJS..169...83L} conducted by Liu et al. (in preparation). Given effective temperature, surface gravity, metallicity, radial velocity, and stellar rotation, this model can predict $H\delta$ and $H\gamma$ line profiles. The radial velocity and other stellar parameters of the spectrum in interest can be determined by comparing to the predicted lines. The typical uncertainty of radial velocity is $\sim5$\,\kms.

The three velocity components in Galactocentric cylindrical coordinates, $V_R$, $V_\phi$, and $V_Z$, are calculated from the radial velocity, distance and proper motions adopting that the motion of the Sun with respect to the local standard of rest (LSR) is (9.58, 10.52, 7.01)\,\kms\ \citep{2015ApJ...809..145T} and the rotational speed of LSR at $V_{\rm LSR}=238$\,\kms\ \citep{2014ApJ.783.130R}. 

\begin{figure}[!ht]
    \centering
    \includegraphics[width=1.0\columnwidth]{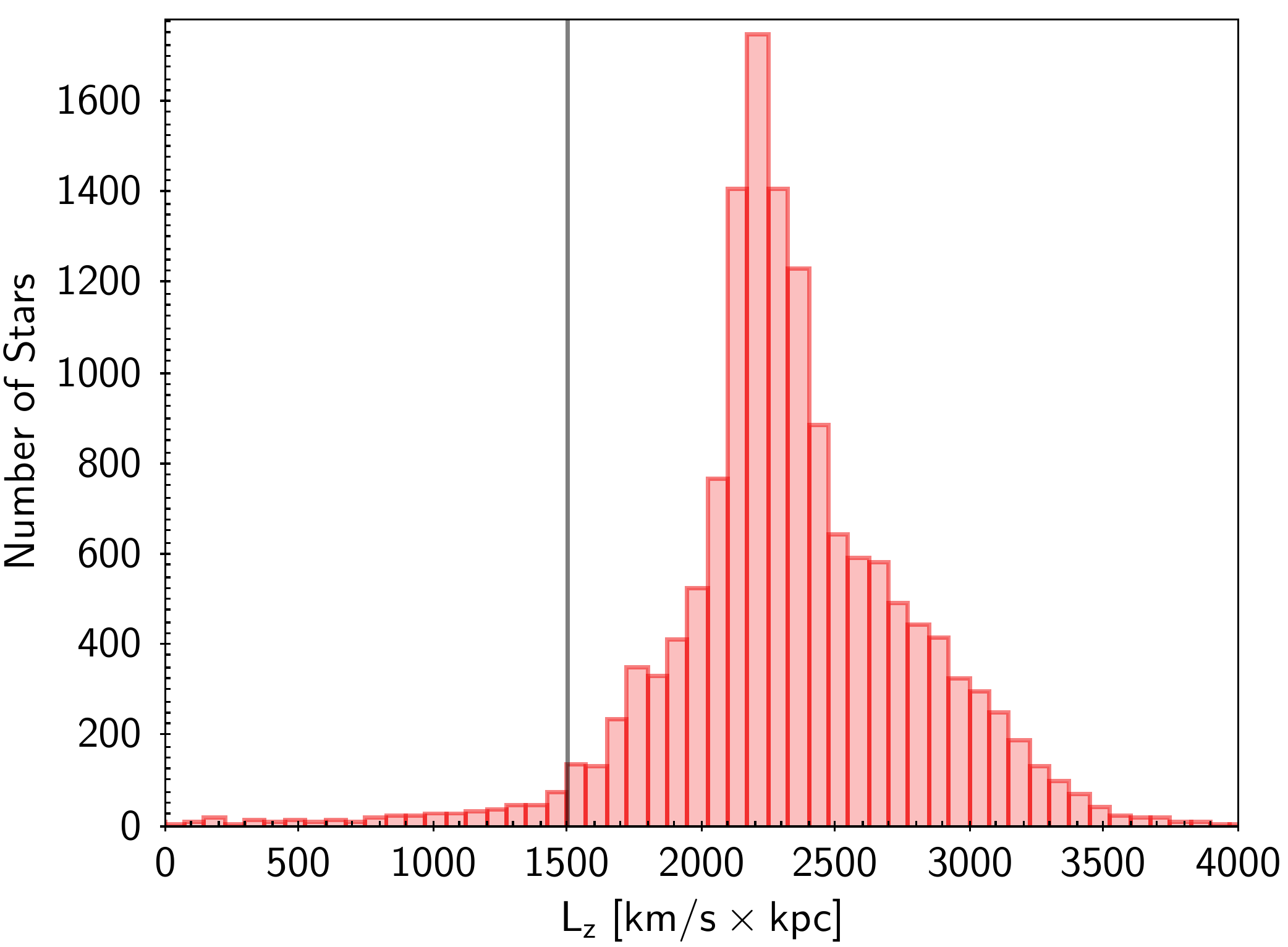}
    \caption{Distribution of $L_z$ of OB star candidates before selection. We exclude possible contamination with $L_z < 1500$\,\kmskpc\, indicated with the black line.}\label{fig.lz_dist}
\end{figure}

Most OB stars are massive stars in the disk and as young as a few tens of millions years, whereas a few low-mass post-AGB and blue horizontal branch stars from the halo population are contaminations. It is not easy to distinguish them from low resolution spectra, but can be identified by locations and kinematics. We only select stars within $|Z|<1.0\text{ kpc}$ and with large enough angular momenta with respect to the $Z$-axis ($L_z$). \Cref{fig.lz_dist} shows the distribution of $L_z$. Only a small fraction of stars have low $L_z$, which may be from the thick disk or the halo. We thus empirically exclude stars with $L_z<1500$\,\kmskpc.

\begin{figure}[!ht]
    \centering
    \includegraphics[width=1.0\columnwidth]{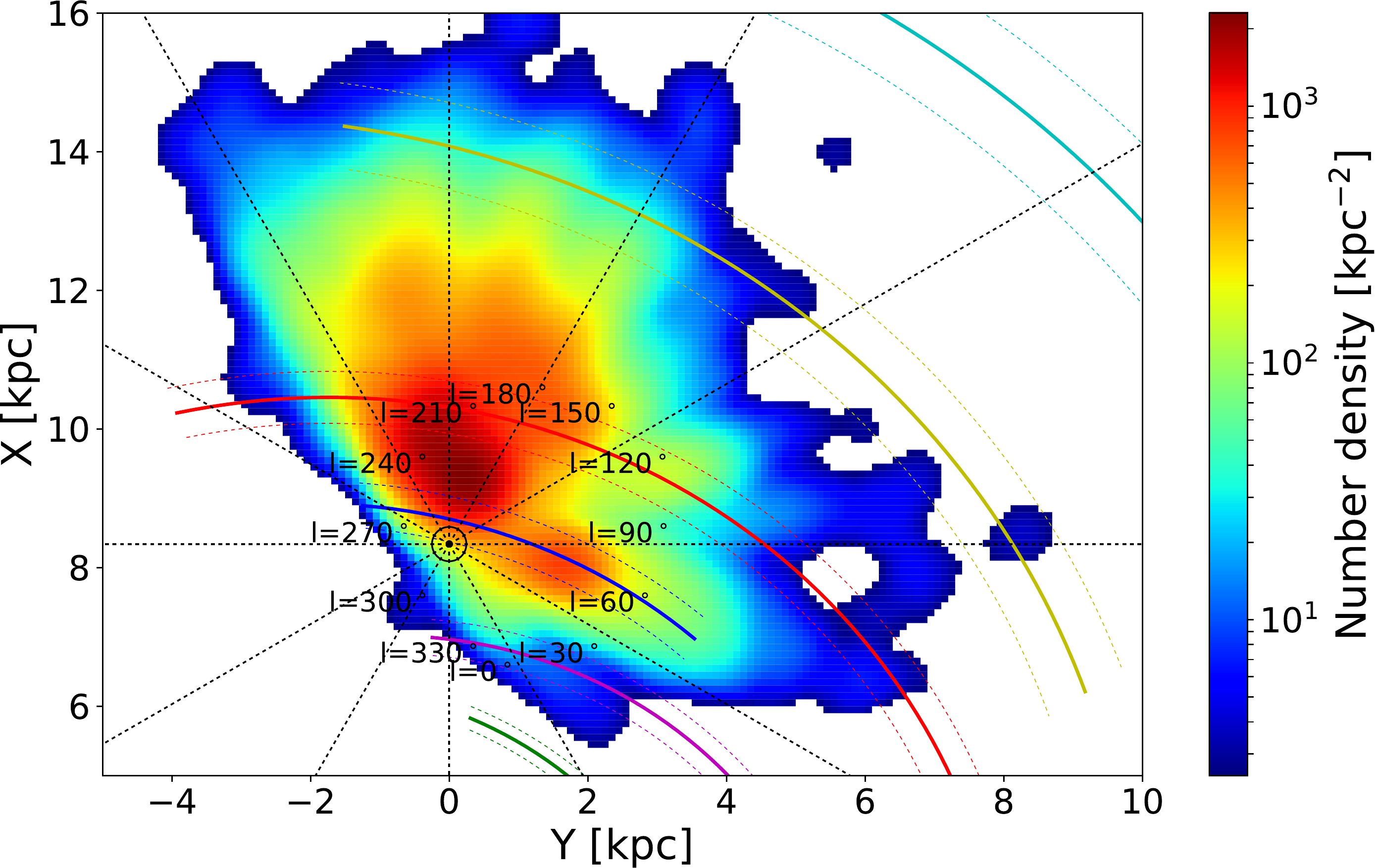}
    \caption{Spatial distribution of the OB stars in the Galactic plane. The Sun marked as $\odot$ is located at ($X=8.34, Y=0$)\,kpc. Spiral structures, i.e. Scutum, Sagittarius, Local, Perseus, Outer, and a new spiral arm, are indicated from the Galactic Center (GC) outwards as solid colored curves. The dotted curves around the solid ones are the uncertainties of the corresponding spiral arms. Their locations are adopted from \citet{2014ApJ.783.130R} and \citet{2015ApJ.798L.27S}. The black dotted lines represent different Galactic longitudes.}\label{fig.obstar_dist}
\end{figure}

We arbitrarily select one spectrum for stars with multiple observations, and obtain 12\,360 OB stars after the above selections. Their spatial distribution in the $X$--$Y$ plane is illustrated in \Cref{fig.obstar_dist}. The galactocentric coordinates are selected such that the Sun is at $X=8.34$\,kpc and $Y=0$\,kpc. The locations of spiral arms are from \citet{2014ApJ.783.130R} and \citet{2015ApJ.798L.27S}. The number density is obtained using the adaptive probability density estimation method with a bi-weight kernel \citep{1999MNRAS.308..731S}. For our sample, the optimal value for the smoothing parameter is $h=0.788$\,kpc.


\section{Results}
\label{sec:res}
\subsection{Ripple patterns in velocity map} 
\begin{figure*}
    \centering
    \subfloat[Radial velocity \VR]{
        \includegraphics[width=0.45\textwidth]{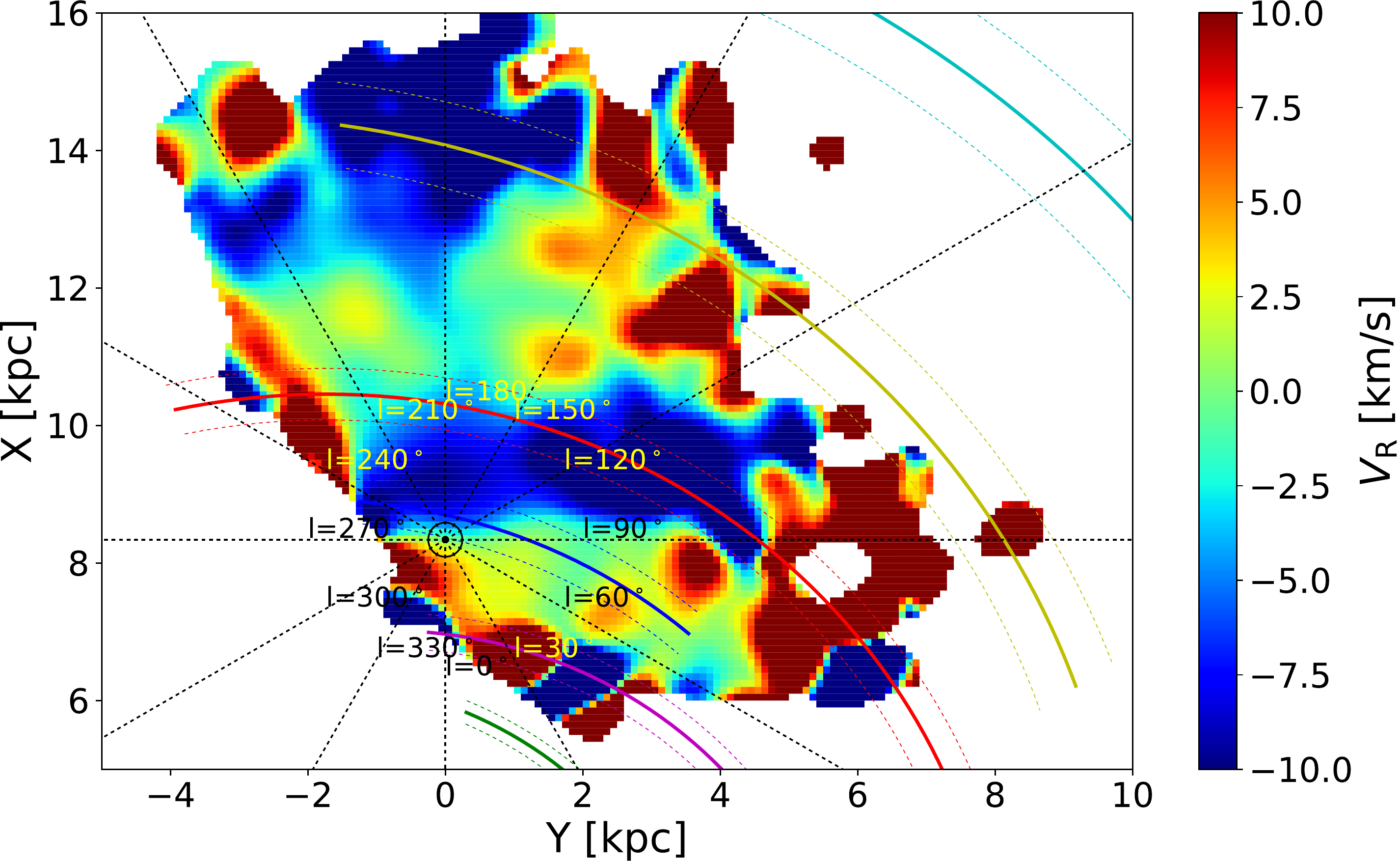}
        \label{fig.vr_x_y}
    }
    \hspace*{\fill}
    \subfloat[Radial velocity uncertainty \EVR]{
        \includegraphics[width=0.45\textwidth]{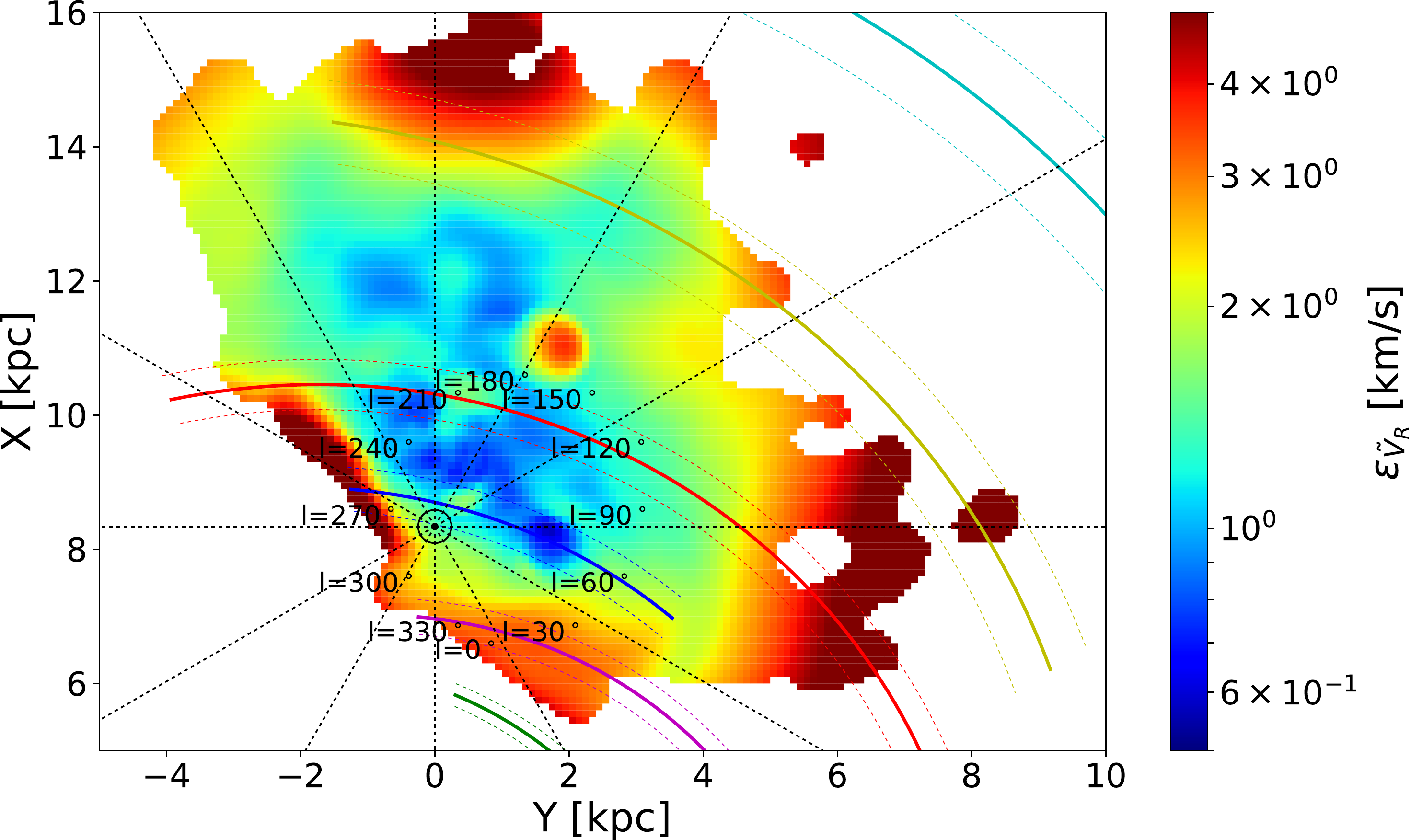}
        \label{fig.evr_x_y}
    }\\[-2ex]
    \subfloat[Azimuthal velocity \dVPHI]{
        \includegraphics[width=0.45\textwidth]{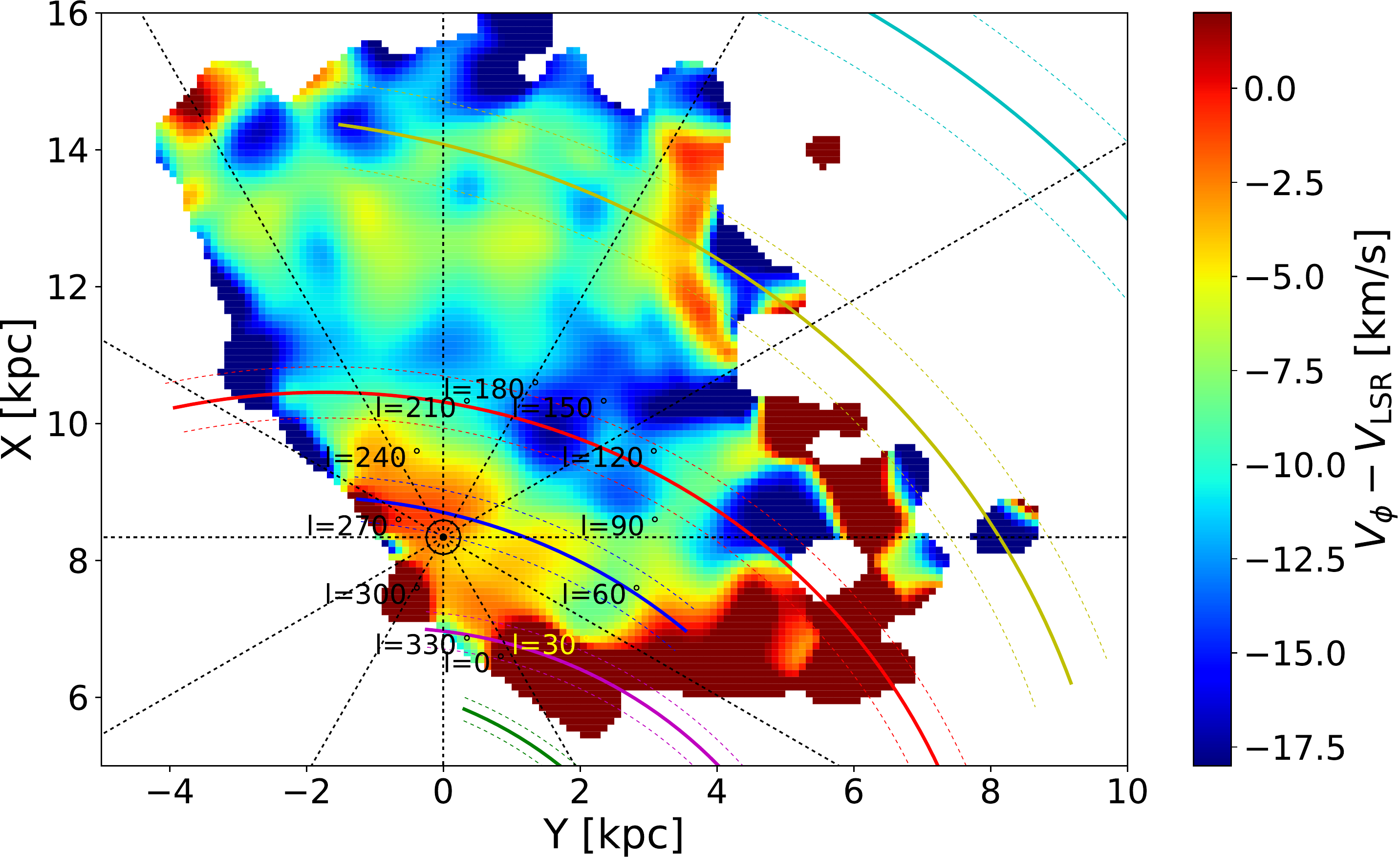}
        \label{fig.vphi_x_y}
    }
    \hspace*{\fill}
    \subfloat[Azimuthal velocity uncertainty \EVPHI]{
        \includegraphics[width=0.45\textwidth]{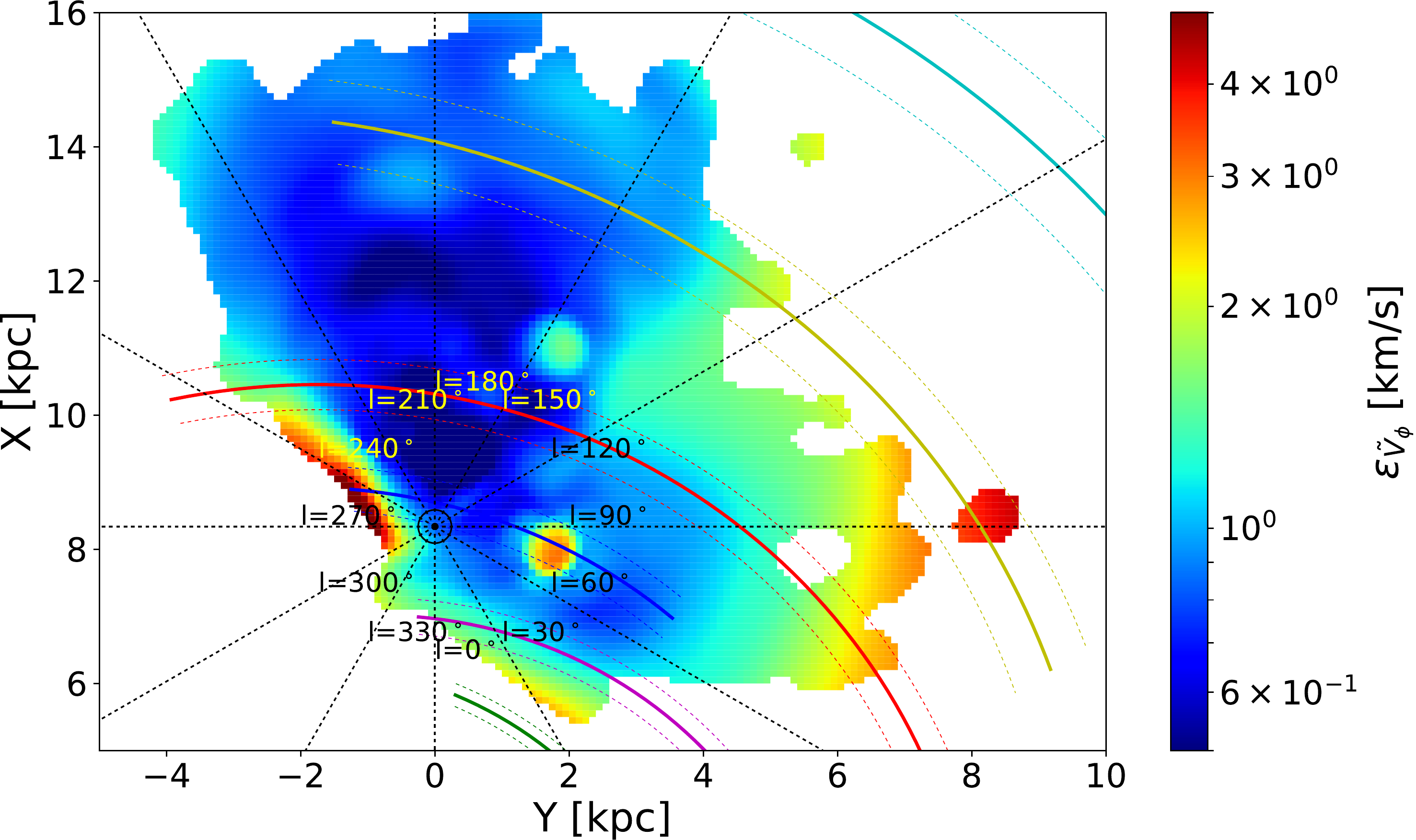}
        \label{fig.evphi_x_y}
    }\\[-2ex]
    \subfloat[Vertical velocity \VZ]{
        \includegraphics[width=0.45\textwidth]{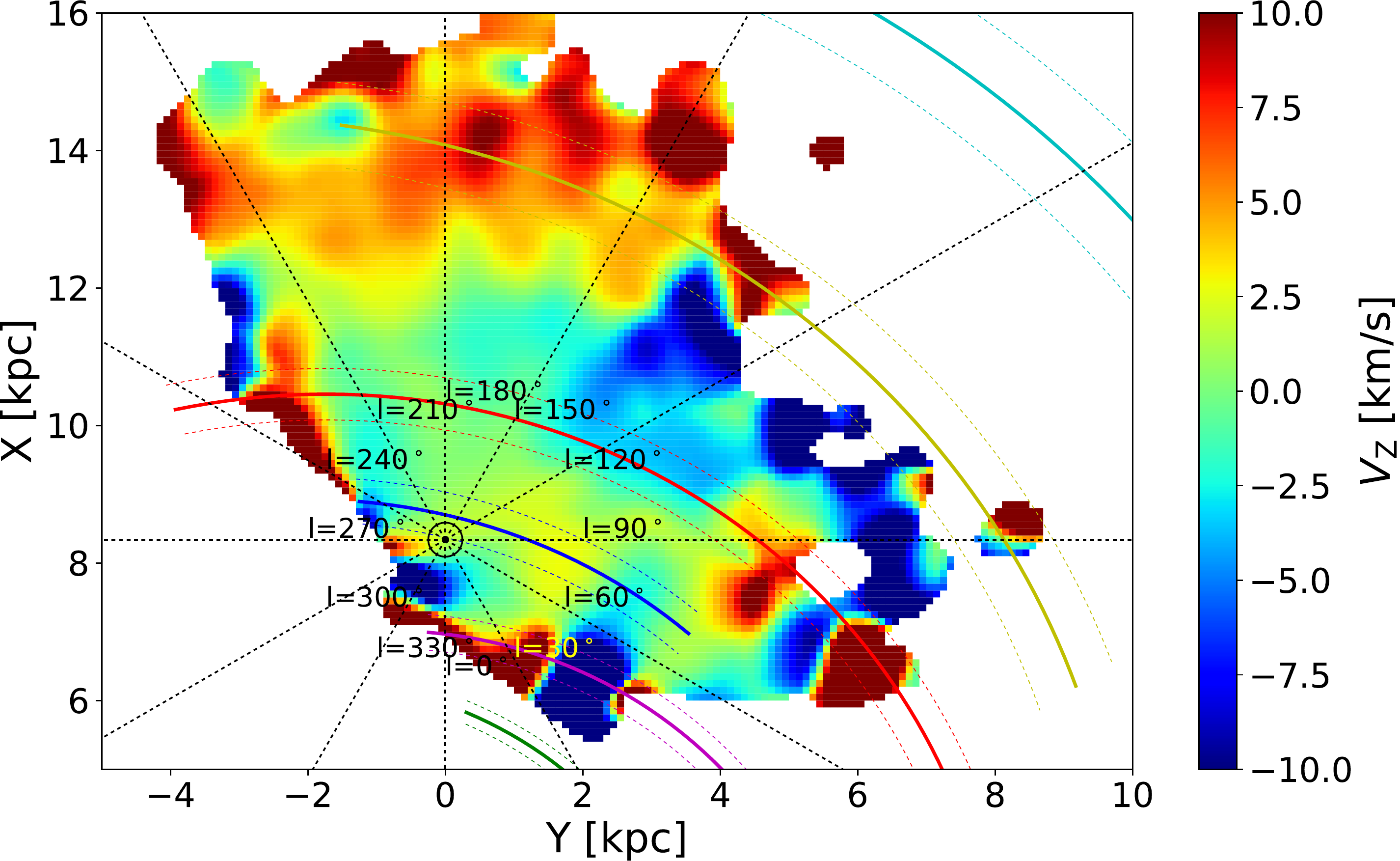}
        \label{fig.vz_x_y}
    }
    \hspace*{\fill}
    \subfloat[Vertical velocity uncertainty \EVZ]{
        \includegraphics[width=0.45\textwidth]{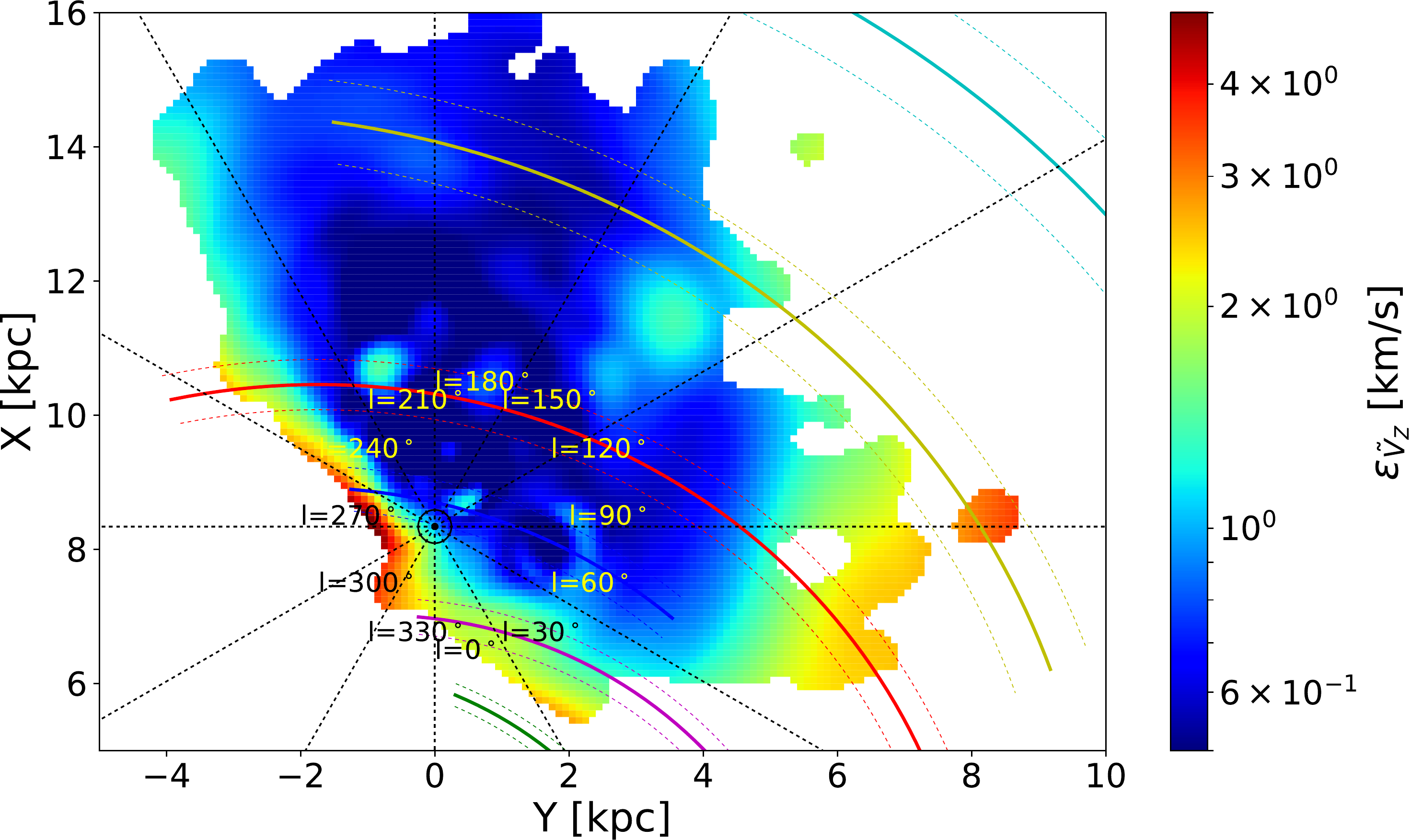}
        \label{fig.evz_x_y}
    }\\[-2ex]
    \subfloat[Radial velocity with $\sigma_d/d<0.1$]{
        \includegraphics[width=0.32\textwidth]{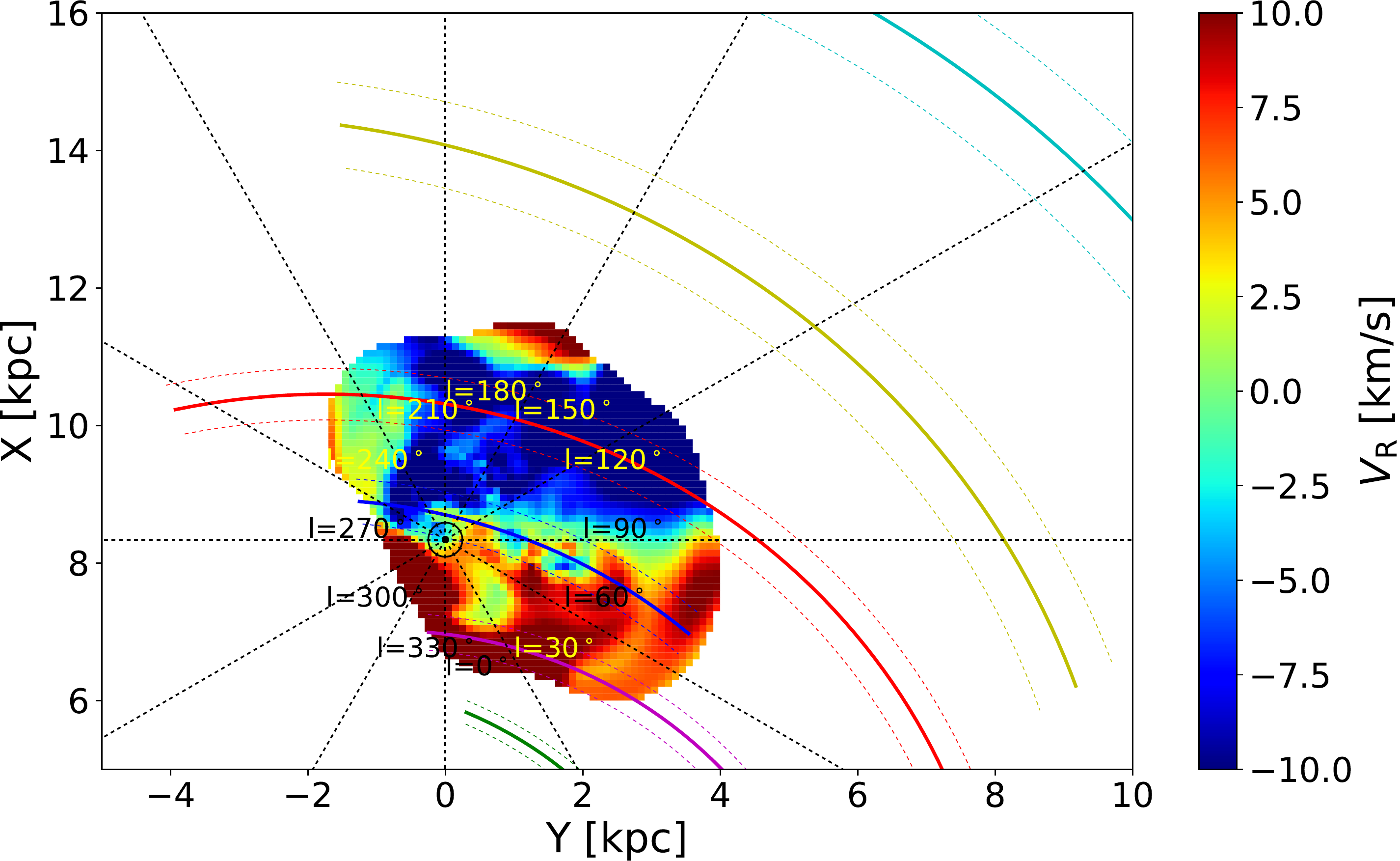}
    }
    \hspace*{\fill}
    \subfloat[Radial velocity with $\sigma_d/d<0.2$]{
        \includegraphics[width=0.32\textwidth]{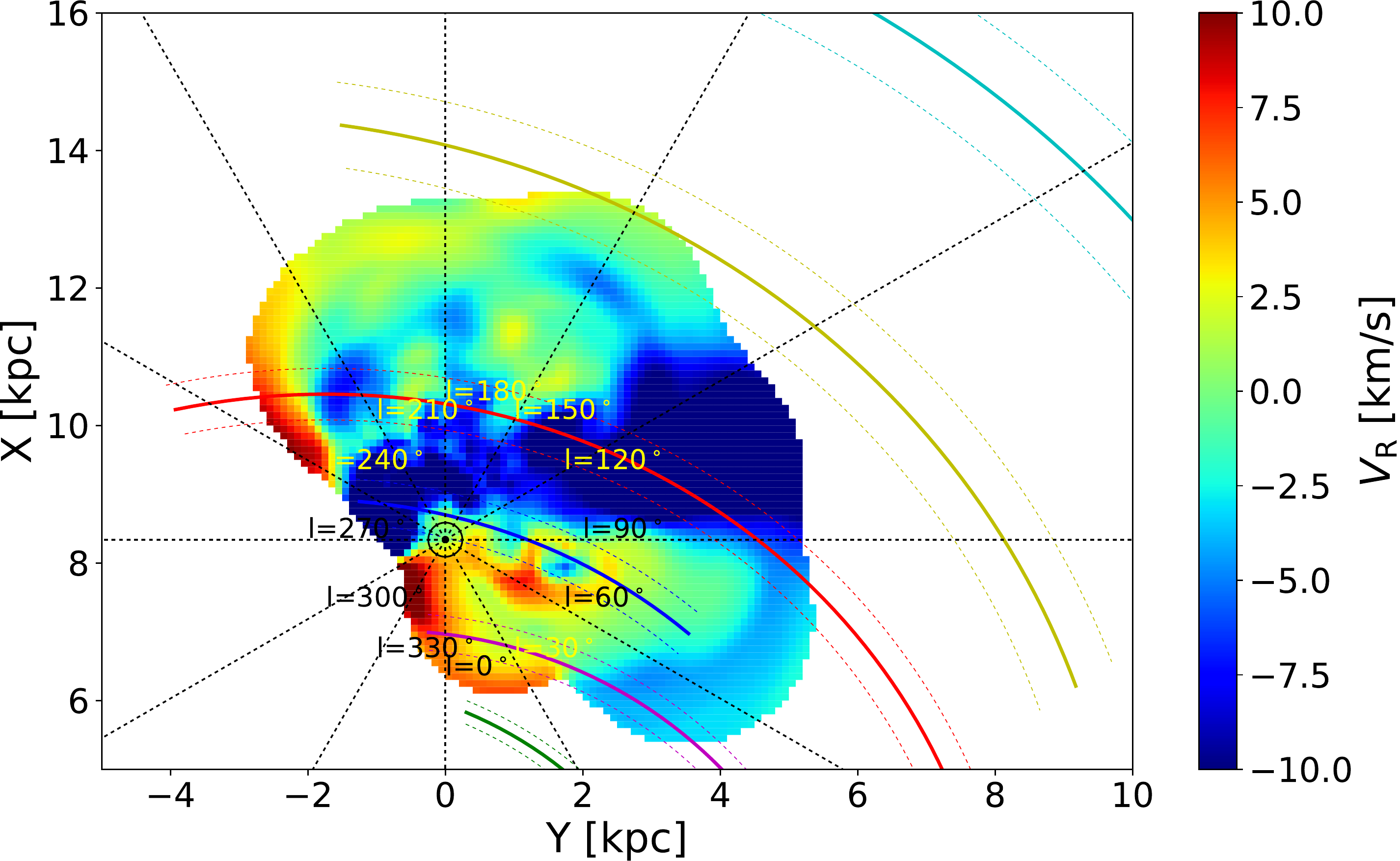}
    }
    \hspace*{\fill}
    \subfloat[Radial velocity with $\sigma_d/d<0.5$]{
        \includegraphics[width=0.32\textwidth]{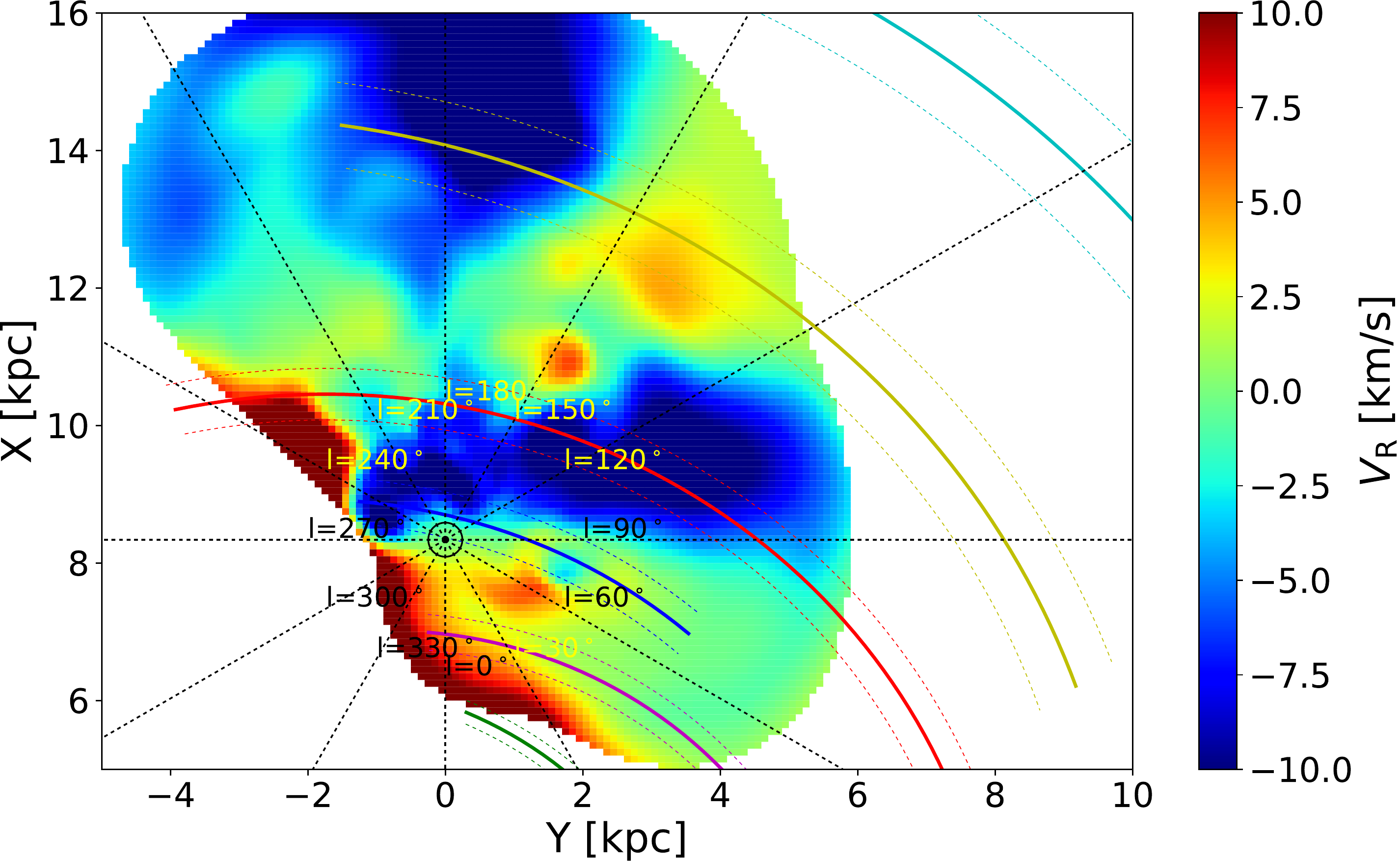}
    }
    \caption{Figure (a), (c) and (e) are the maps of velocity components, \VR, \VPHI, and \VZ\ from top to bottom, respectively, Figure (b), (d) and (f) are their respective uncertainties. The bottom row tests the \VR\ maps of the stars with $\sigma_d/d<0.1$, 0.2 and 0.5 in (g), (h), and (i) respectively. The Sun is located at ($X=8.34$, $Y=0$)\,kpc and is marked as $\odot$. The solid curves indicate the spiral structures as in \Cref{fig.obstar_dist}. The dotted lines represent different Galactic longitudes. }\label{fig.v_x_y}
\end{figure*}

\Cref{fig.v_x_y}(a), (c) and (e) respectively show the velocity fields of \VR, \VPHI\ and \VZ\ estimated with the same technique as for the number density, while panels (b), (d) and (f) show the uncertainties of the corresponding velocity fields estimated from bootstrapping. We arbitrarily selected 80\% stars as a subsample and added corresponding Gaussian noise to their velocities. The adaptive probability density estimation was rebuilt for this subsample to derive the velocity fields. The process was repeated 1000 times, and the standard deviations at each point in the $X$--$Y$ plane are calculated and taken as uncertainties. Note that they are different from the velocity dispersions by definition.

\begin{figure}[ht]
    \includegraphics[width=\columnwidth]{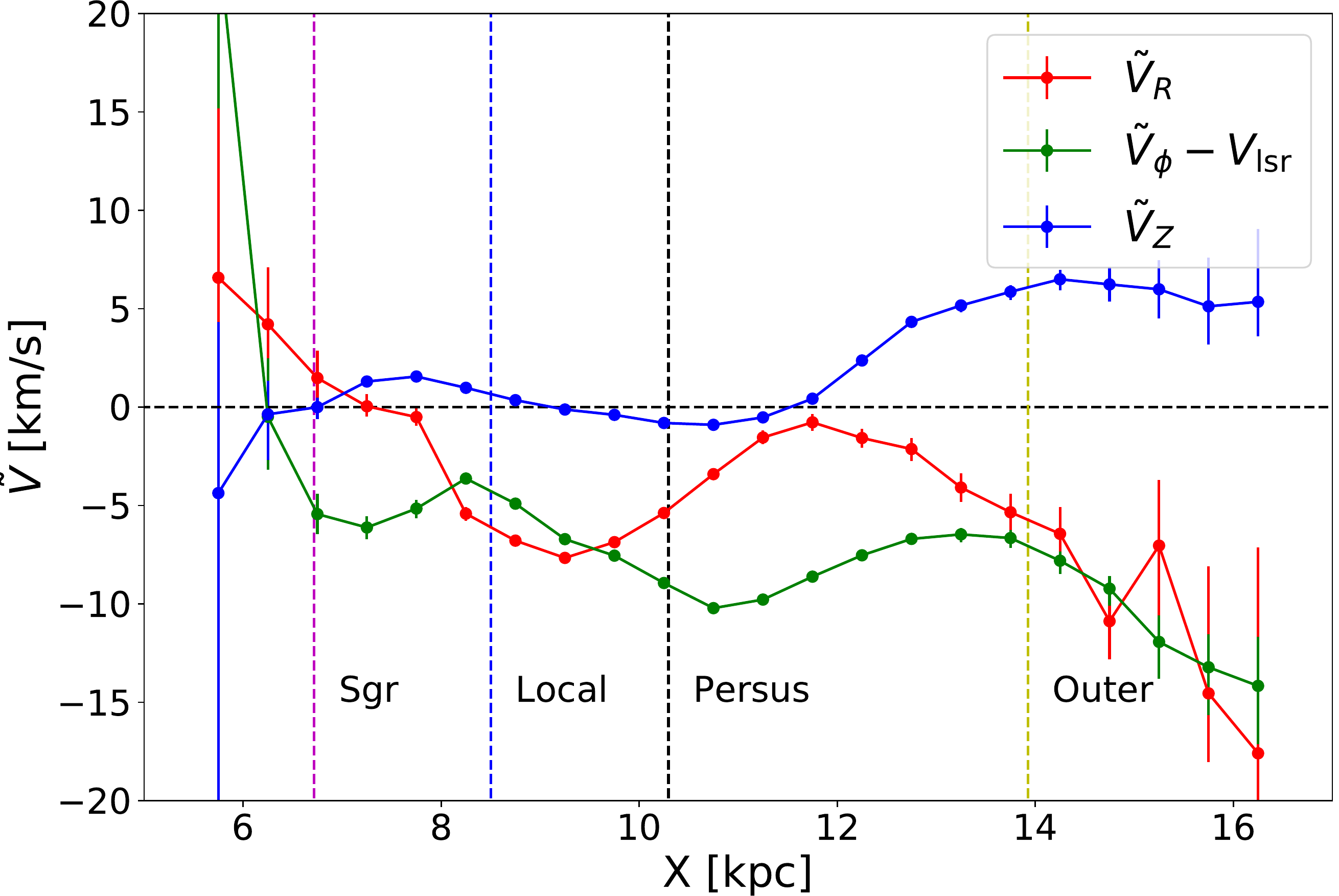}
    \caption{Average velocities with respect to $X$ with stars in $-2.5<Y<2.5$ kpc. The average positions of spiral arms are indicated with dotted vertical lines.}\label{fig.vx}
\end{figure}

We integrated the three velocity fields within $-2.5<Y<2.5$ kpc and plotted the average velocity with respect to $X$ in \Cref{fig.vx} to better illustrate the radial variation of velocities. The uncertainties are estimated with the same methods as mentioned above.

\subsubsection{Asymmetry in $V_R$}
A few local ripple-like patterns are identified in the map of \VR\ \Cref{fig.vr_x_y}. That is, \VR\ displays with two negative-velocity strips (blue) with a zero-value strip in between. At $X\sim9$\,kpc, \VR\ declines to $\sim-8$ km/s, and increases back to $\sim0$ at around $X=12$\,kpc, and finally drops to below $-10$\,\kms\ when $X>15$\,kpc. Both negative \VR\ features are detected with a $>5\sigma$ confidence level according to the map of \EVR\ shown in \Cref{fig.evr_x_y}. These features are better shown in \Cref{fig.vx} with the red lines. 

Compared to the spiral arms, the ripple patterns are not aligned with them in the $X$--$Y$ plane. The near negative \VR\ strip overlaps with the local arm at ($X\sim9$, $Y\sim0$)\,kpc. It does not follow the arm to ($X=7.5$, $Y=3.5$)\,kpc, but extends along $X\sim9$\,kpc and overlaps with part of the Perseus arm at $Y\sim3$\,kpc. At $Y>3$\,kpc, it goes beyond the Perseus arm. Similarly, the farther negative \VR\ strip is located at larger radius than the Outer arm. In the bottom row of \Cref{fig.v_x_y}, we tested the robustness of the ripple as a function of distance uncertainty. We cut the data with $\sigma_d/d<0.1$, 0.2, and 0.5. The 9\,kpc \VR\ strip exists in all three cases and show similar trend for the case with $\sigma_d/d<0.2$. This strongly indicates that the pattern is real.

The strip centered at $X=9$ kpc is consistent with previous discoveries \citep{2011MNRAS.412.2026S,2013ApJ.777L.5C, 2017RAA.17.114T, 2017ApJ...835L..18L,2018IAUS.334.109L, 2018A&A.616A.11G}. \citet{2012MNRAS.425.2335S} and \citet{2018A&A.616A.11G} suggested that this asymmetric radial velocity is produced by the perturbation of nearby spiral structures. Theoretical studies \citep{2014MNRAS.443L.1D,2014MNRAS.440.2564F} found that if this is the case, then the spatial distribution of the asymmetric \VR\ should be naturally correlated with the coherent spiral structures. Therefore, the uncorrelated spatial distribution between the negative \VR\ strip and the local or Perseus arm shows that the perturbation of the spiral arms is not the main driver of the ripples.

Compared to \cite{2016ApJ...823....4D}, we find the ripple patterns are similar to their simulations (see the top-right plot in their Figure 4), which demonstrate a perturbed disk by an interaction with a satellite.

Note that the strip at $X>15$\,kpc approaches the regime that the parallax from \gaia\ is less precise. The larger uncertainties of parallax and the uncertainties of proper motions have been considered in the estimation of the uncertainty of \VR\ shown in \Cref{fig.evr_x_y}. According to this figure, the typical error of the median radial velocity at $X\sim15$\,kpc is around 3-4\,\kms. The smaller values of error are mostly due to two reasons. First, the radial velocity is mostly contributed by the line-of-sight velocity derived from the LAMOST spectra, which is independent of parallax. Second, the large number of OB stars detected reduces the uncertainty of the velocity, even though the typical error of the line-of-sight velocity is $\sim5$\,\kms. Therefore, the strip shown at $X>15$\,kpc is real. Similar to the strip displayed at $X\sim9$\,kpc, the misalignment of the farther strip with the Outer arm implies that the spiral arms are not responsible for the ripples.

\subsubsection{Asymmetry in \dVPHI\ and \VZ}
Similar ripple patterns are also seen in \dVPHI, as shown in \Cref{fig.vphi_x_y} and \Cref{fig.vx} (green line). Unlike \VR, \dVPHI\ shows three dips at around $X=7$, $11$\,kpc and $>15$\,kpc. We can conclude from \Cref{fig.vx} that the ripple of \dVPHI\ has roughly $1/4$ phase difference with the ripple of \VR. It clearly shows that the dip of \dVPHI\ at $X\sim11$\,kpc is located between the dip of \VR\ at $\sim9$\,kpc and the peak of \VR\ at $\sim12$\,kpc. Meanwhile, the local peak of \dVPHI\ at around $13$\,kpc is beyond the peak of \VR\ located at $\sim12$\,kpc. 

Indeed, if the disturber, internal (e.g. spiral structure) or external (e.g. a dwarf galaxy), introduces an in-plane gravitational force to the stars, the decomposition of the perturbation force along the radial and azimuthal direction would naturally induce additional radial and azimuthal motions with $1/4$ phase difference to the stars.

\citet{2017ApJ...835L..18L} showed a similar azimuthal velocity map with young stars. However, because their young F-type stars only covers less than 1\,kpc around the Sun, they did not find that the velocity pattern is unrelated to the local arm at larger distance. 

The ripple pattern is not clearly displayed in the distribution of the vertical velocity, but other interesting features are seen in the \VZ\ map. A region with a relatively high upward motion ($>5$\,\kms) beyond $X=13$\,kpc is discovered with $3\sim 5\sigma$ confidence level (see \Cref{fig.vz_x_y} and more clearly in \Cref{fig.vx}).

\citet{2017IAUS.321.6L} discovered similar vertical motion using red clump stars along the Galactic anti-center direction and attributed it to the Galactic warp. \citet{2018MNRAS.481L.21P} reported a gradient of similar scale at 8-14 kpc. Furthermore, similar to their findings, the upward vertical velocity also shows a larger amplitude in the second quadrant of the disk plane, i.e. $Y>0$ in our coordinates. This implies that the line-of-node of the warp may be located in the third quadrant.

\subsubsection{A substructure along $l=90^\circ$ ?}
Beside the features discussed above, a suspected substructure is found in the velocity map along the Galactic longitude of $l=90^\circ$ beyond $Y>5$\,kpc along the longitude. The OB stars in this region show \VR$>+10$\,\kms, \dVPHI$<-15$\,\kms, and \VZ$<-5$\,\kms. In other word, this group of stars is moving away from the Galactic center and downward to the south with slower azimuthal velocity. The feature is close to the edge of the detection of the sample. Therefore, it is not clear whether this is real.

\subsection{Spiral feature in the phase space}
\begin{figure*}[ht]
    \centering
    \subfloat[Full sample]{
        \includegraphics[width=0.45\textwidth]{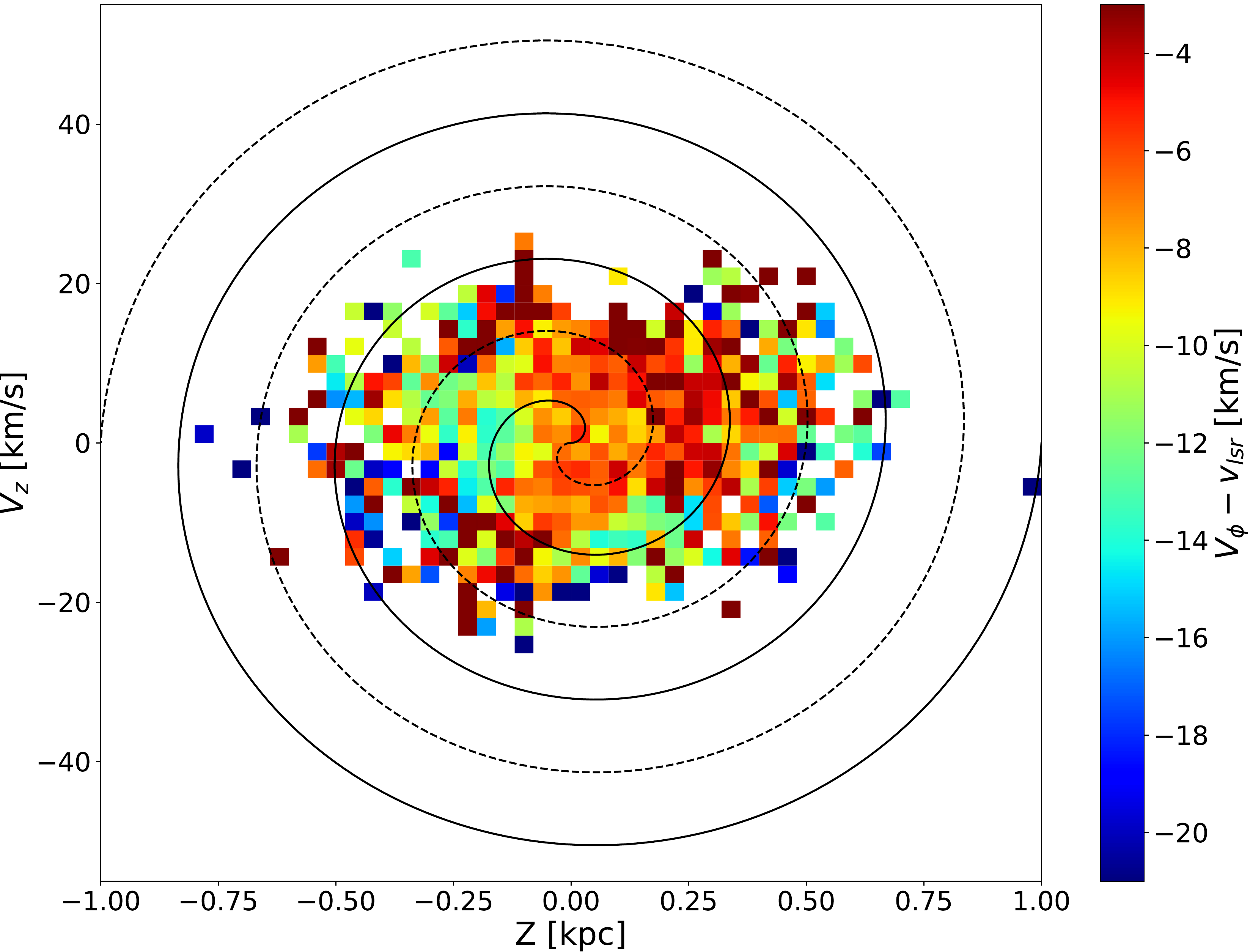}
        \label{fig.pm_full}
    }
    \hspace*{\fill}
    \subfloat[$|R-R_0|<1.0\text{ kpc}$ Cylinder]{
        \includegraphics[width=0.45\textwidth]{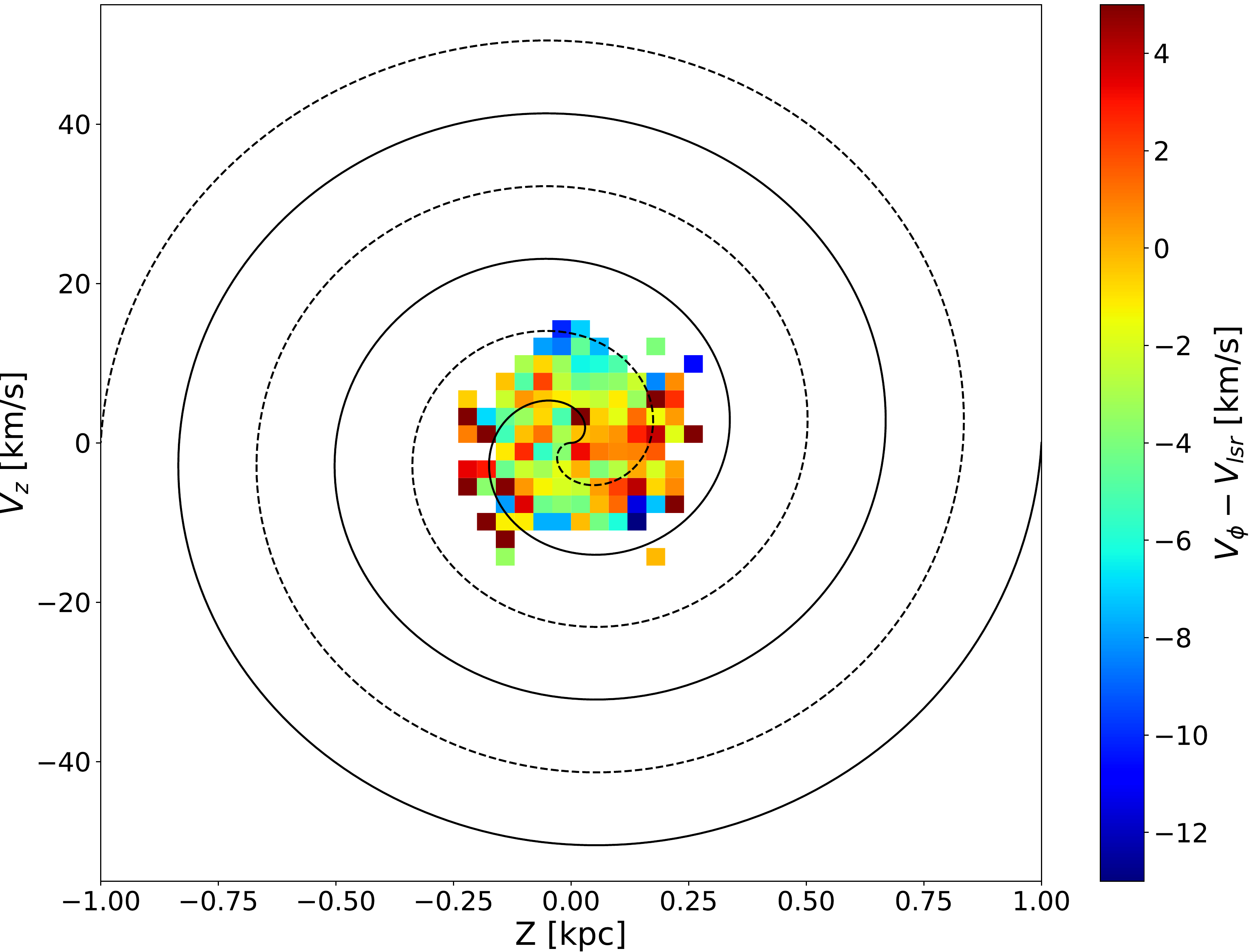}
        \label{fig.pm_part}
    }
    \caption{\dVPHI\ in the $Z$--$V_Z$ phase space. The solid and dashed lines indicate the spiral reported by \citet{2018arXiv180902658B}, with the solid line as the high peak and the dashed line as the valley between peaks. The left panel is for the full sample, while the right panel is for stars witin 1\,kpc of the Sun.}
\end{figure*}

Following \citet{2018ApJ.865L.19T}, who tried to find the spiral feature in the phase space for younger stars so that the time of the occurrence of the perturbation can be better constrained, we examine the distribution of \VPHI\ in the $Z$--$V_Z$ phase space with our samples. 

\Cref{fig.pm_full} shows the map of the median \VPHI\ with subtraction of $V_{\rm LSR}$ in the $Z$--$V_Z$ plane for all OB stars, while \Cref{fig.pm_part} displays the similar map but only with stars in the same cylinder selected by \citet{2018arXiv180902658B}. We see that \dVPHI\ varies in the $Z$--$V_Z$ plane with amplitude of $4\sim8$\,\kms. The lowest \dVPHI\ ($-2\sim-4$ km/s) appears at ($Z\sim-0.25$\,kpc, \VZ$\sim0$\,\kms), and the highest values ($2\sim4$\,\kms) appear at roughly the mirroring position of the lowest region with respect to $Z=0$. However, it does not show a clear spiral pattern as seen in other works. This is reasonable since the OB stars are too young to finish even one orbital period around the Galactic center.

\section{Discussions and conclusions}\label{sec:conc}
It seems that the ripple patterns in the map of in-plane velocities is the result of a perturbation. In principle, the disturber could be classified as internal, such as spiral structures, giant molecular clouds or a central rotating bar, or external, such as pass-by dwarf galaxies or dark matter sub-halos. In the next paragraphs we attempt to briefly discuss these possibilities in turn.

As discussed in the previous section, it is clear that the ripple patterns shown in the map of \VR\ and \VPHI\ do not follow the present-day gaseous spiral arms. Thus, they are the result of perturbations induced by current spiral structures according to \citet{2012MNRAS.425.2335S}. However, it is noted that they assume a steady spiral structure in their simulation, while the origin of the spiral structures is far from clear \citep{2018ApJ...853L..23B, 2018AJ....156..248T, 2018arXiv181003325S}. Therefore, our results may not clearly exclude the perturbation of spiral arms originated from other mechanism but only rule out a steady density wave spiral structure.

\citet{2018IAUS.334.109L} revealed that a bar with pattern speed of 60\,\kmskpc\ is also capable of producing the dip in \VR\ centered at $X\sim9$\,kpc through a test particle simulation, but could not produce the farther decline beyond $13$ kpc. Thus, the bar alone is not the reason.

Giant molecular clouds may also be potential disturbers and can induce spiral structures according to \citet{2013ApJ...766...34D}. However, this scenario can not be investigated until more simulations about how giant molecular clouds affect the stellar kinematics are performed.

Considering the external disturbers, many recent works in Section~\ref{sec:intro} argued that Sagittarius is one of the possible disturbers. However, it is unlikely that these OB stars were directly affected due to their young age. Instead, the encounter might affect the gaseous disk, which may eventually affect the orbits of the OB stars formed from it. As pointed out by \citet{2018MNRAS.481.1501B} and \citet{2018arXiv180902658B}, the pass-by satellite attracts the disk stars both in vertical and in-plane direction. The in-plane perturbation may raise complicated spiral structures with rich ripple-like in-plane velocities as seen in the simulation by \citet{2018arXiv180902658B} (See their Figure 25). Unlike the stellar disk, dissipation may play an important role in the gaseous disk.

As a qualitative estimate, the speed of sound can be used to obtain an approximate time timescale of dissipation. The cold neutral medium (CNM) has a typical temperature of $T=100\text{ K}$ and hydrogen inside is in atomic state \citep{2001RevModPhys.73.1031}. Thus, we can treat CNM as ideal gas and estimate the speed of sound in adiabatic and isothermal processes.
\begin{equation}\label{eq:c_adi}
    V_{\rm ad} = \sqrt{\frac{\gamma kT}{m}}=1.2\text{ km/s},
\end{equation}
\begin{equation}\label{eq:c_iso}
    V_{\rm iso} = \sqrt{\frac{kT}{m}}=0.9\text{ km/s},
\end{equation}
where $\gamma=5/3$ for hydrogen in atomic state, $k$ is the Boltzmann constant, and $m$ is the mass of the hydrogen atom. Thus, $V_{\rm sound}=1\text{ km/s}$ is a reasonable approximation for the speed of sound in the CNM.

The scale height of HI region at the position of the Sun is $\sim0.15$\,kpc \citep{2009ARA&A.47.27K}. Therefore, a vertical perturbation would dissipate in a time scale of
\begin{equation}\label{eq:vert}
    \tau_{\rm vertical} = \frac{H_{\rm vertical}}{V_{\rm sound}}=\frac{150\text{ pc}}{1\text{ km/s}}\approx150\text{ Myr}.
\end{equation}
Considering that the encounter was $\sim500$\,Myrs ago, it is possible that a great proportion of the vertical impact in the gaseous disk has been dissipated. This can explain the lack of spiral features in the vertical phase space shown in \Cref{fig.pm_full}. 

On the other hand, the in-plane perturbation due to the encountering satellite can induce spiral structures in the gaseous disk, similar to the stellar disk. However, due to dissipation of the gas, the spiral structures may be finally damped as mentioned in Section 6.4 of \citet{2008gady.book.....B}. A very rough estimate of the time scale of dissipation in the in-plane direction can be made using the size of the gaseous disk and the speed of sound. Considering the scale length of gaseous disk is $\sim3.15$\,kpc \citep{2009ARA&A.47.27K}, the timescale of dissipation can be approximated as 
\begin{equation}\label{eq:horiz}
    \tau_{\rm H} = \frac{H_{\rm r}}{V_{\rm sound}}=\frac{3.15\text{ kpc}}{1\text{ km/s}}\approx3.1\text{ Gyr}.
\end{equation}
 
This means that although a perturbation to the disk occurred a few hundred Myrs ago, the marks of the in-plane perturbation sculpted in the gaseous disk may later be inherited by the new born OB stars from these gas clouds.

An alternative scenario is that another dark matter sub-halo disturbed the disk a few tens of Myrs ago and imprinted the ripple pattern in these young stars. If this is the case, similar patterns should also be seen in relatively old stellar populations.

In summary, we explored the kinematic structure of the Milky Way galaxy with OB stars. While they do not show a clear indication of the phase-mixing spiral in the $Z$--$V_Z$ plane, we found clear ripple patterns in the in-plane map of \VR\ and \VPHI. The roughly $1/4$ phase difference between the ripple features in \VR\ and \VPHI\ implies that these patterns should be a result of an in-plane perturbation. We considered a few possible disturbers, including internal and external origins, and find that neither of them can be the exclusive reason for the ripple structure, although perturbation of stationary spiral structures can be ruled out since the ripples are not aligned with the known spiral structures.

\acknowledgements
We appreciate the anonymous referee for his/her helpful suggestions which improved the paper. We thank Youjun Lu, Zhao-Yu Li, Juntai Shen, and Elena D'Onghia for helpful discussions and comments. This work is supported by the National Key Basic Research and Development Program of China No. 2018YFA0404501 and NFSC under grants 11821303, 11390372, 11761131004 (SM), 11333003 (CL and SM), 11873057 (CL) and 11773009 (WC). Guoshoujing Telescope (the Large Sky Area Multi-Object Fiber Spectroscopic Telescope LAMOST) is a National Major Scientific Project built by the Chinese Academy of Sciences. Funding for the project has been provided by the National Development and Reform Commission. LAMOST is operated and managed by the National Astronomical Observatories, Chinese Academy of Sciences. This work has made use of data from the European Space Agency (ESA) mission {\it Gaia} (\url{https://www.cosmos.esa.int/gaia}), processed by the {\it Gaia} Data Processing and Analysis Consortium (DPAC, \url{https://www.cosmos.esa.int/web/gaia/dpac/consortium}). Funding for the DPAC has been provided by national institutions, in particular the institutions participating in the {\it Gaia} Multilateral Agreement.


\end{document}